\documentclass[journal=ancac3,manuscript=article]{achemso}
\usepackage{graphicx}
\usepackage{hyperref} 
\usepackage{amssymb}
\usepackage{color}
\usepackage{filecontents}

\author{Fausto Martelli}
\affiliation{IBM Research Europe, Hartree Centre, Daresbury, WA4 4AD, United Kingdom}
\email{fausto.martelli@ibm.com}
\author{Jason Crain}
\affiliation{IBM Research Europe, Hartree Centre, Daresbury, WA4 4AD, United Kingdom}
\alsoaffiliation{Department of Biochemistry, University of Oxford, South Parks Road, Oxford, OX1 3QU, United Kingdom}
\author{Giancarlo Franzese}
\affiliation{Secci\'o de
 F\'isica Estad\'istica i Interdisciplin\`aria--Departament de F\'{i}sica de la Mat\`{e}ria Condensada, Universitat de Barcelona, 
 \& Institut de Nanoci\`{e}ncia i Nanotecnologia (IN2UB), Universitat de Barcelona, C. Mart\'{i} i Franqu\`{e}s 1, 08028 Barcelona, Spain}

\title{Network Topology in Water Nanoconfined between Phospholipid Membranes}

\begin{document}

\begin{abstract}
Water provides the driving force for the assembly and stability of many cellular components. Despite its impact on biological functions, a nanoscale understanding of the relationship
between its structure and dynamics under soft confinement has remained elusive. 
As expected, water in contact with biological membranes recovers its bulk density and dynamics at $\sim 1$ nm  from phospholipid headgroups but surprisingly enhances its intermediate-range order (IRO) over a distance, at least, twice as large.  
Here, we explore how the IRO is related to the water's hydrogen bond network (HBN) and its coordination defects. We characterize the increased IRO by an alteration of the HBN up 
to more than eight coordination shells of hydration water. The  HBN analysis emphasizes the existence of a bound-unbound water interface at $\sim 0.8$ nm from the membrane. The unbound water has a distribution of defects intermediate between bound and bulk water, but with density and dynamics similar to bulk, while bound water has  reduced thermal energy and much more HBN defects than low-temperature water.
This observation could be fundamental for developing nanoscale models of biological interactions and for understanding how alteration of the water structure and topology, for example, due to changes in extracellular ions concentration, could affect diseases and signaling. More generally, it gives us a different perspective to study nanoconfined water.
\end{abstract}

                             
\maketitle

It has been long recognized that the structure and function of biological membranes are largely determined by the properties of hydration water, \textit{i.e.}, of the water in contact with the membrane~\cite{Berkowitz_chemrev2006}. Indeed, the presence of water strongly influences membrane stability, fluidity, and phase behavior, thereby affecting membrane function and properties. Also, hydration water mediates the interactions of biological membranes with other biomolecules and with ions~\cite{bookBiomembranes2011, bookBioMembrane2019}.

Biological membranes are composed of a large number of components, including proteins, cholesterol, glycolipids, and ion channels, among others, but their framework is provided by phospholipid molecules that self-assemble into bilayers driven by the hydrophobic effect~\cite{Fitter_JPhysChemB1999, bookBiomembranes2011, bookBioMembrane2019, Trapp_JCP2010,Wassall_BiophysJ1996,Righini_PRL2007,Zhao_Fayer_JACS2008, Tielrooij_BiophysJ2009,Hua_CPC2015,Rog_ChemPhysLett2002,Bhide_JCP2005,Berkowitz_chemrev2006,Hansen_PRL2013,               Zhang_Berkowitz_JPhysChemB2009,Gruenbaum_JChemPhys_2011,Calero_Mat2016}. To decrease the interfacial free-energy, polar head groups form contacts with water, while the apolar hydrocarbon tails minimize exposure to water forming extended bilayers. Water is abundant in the interfacial region of bilayers (lipid headgroups), establishing strong hydrogen bonds (HBs) with the membrane. As a result of these strong interactions, the orientational and translational dynamics of interfacial water is markedly slowed down~\cite{Righini_PRL2007,Zhao_Fayer_JACS2008, Tielrooij_BiophysJ2009,Zhang_Berkowitz_JPhysChemB2009, Gruenbaum_JChemPhys_2011,Calero_Mat2016,Wassall_BiophysJ1996}. Such slowing down has been observed also in water in contact with proteins and sugars~\cite{camisasca_2018,iorio_2019,iorio_2019_2,iorio_2019_3,iorio_2020}.

Recently we found an increase in the structural order at the intermediate range when the dynamics of
water confined by phospholipid membranes slows down. This intermediate range order (IRO)
 propagates as far as (at least) $\sim2.4$~nm from the membrane surface~\cite{martelli_fop}, a larger distance than previously calculated using other observables such as density and dynamical properties. 
We recovered  
 water's bulk density and dynamical properties at a distance of $\sim1.2$~nm from the membrane surface~\cite{martelli_fop,martelli_chapter}. Nonetheless, water is a complex network-forming material, with a directional HBs network (HBN), the topology of which is correlated to anomalous behavior, as we showed recently~\cite{martelli_unravelling_2019}. Therefore, understanding how the HBN of water is affected by the interactions with the membrane is of primary importance in understanding biological properties at the molecular scale also in conditions in which a biological membrane interacts with alien components such as, \textit{e.g.}, viruses.

\begin{figure}
  \begin{center}
   \includegraphics[scale=.50]{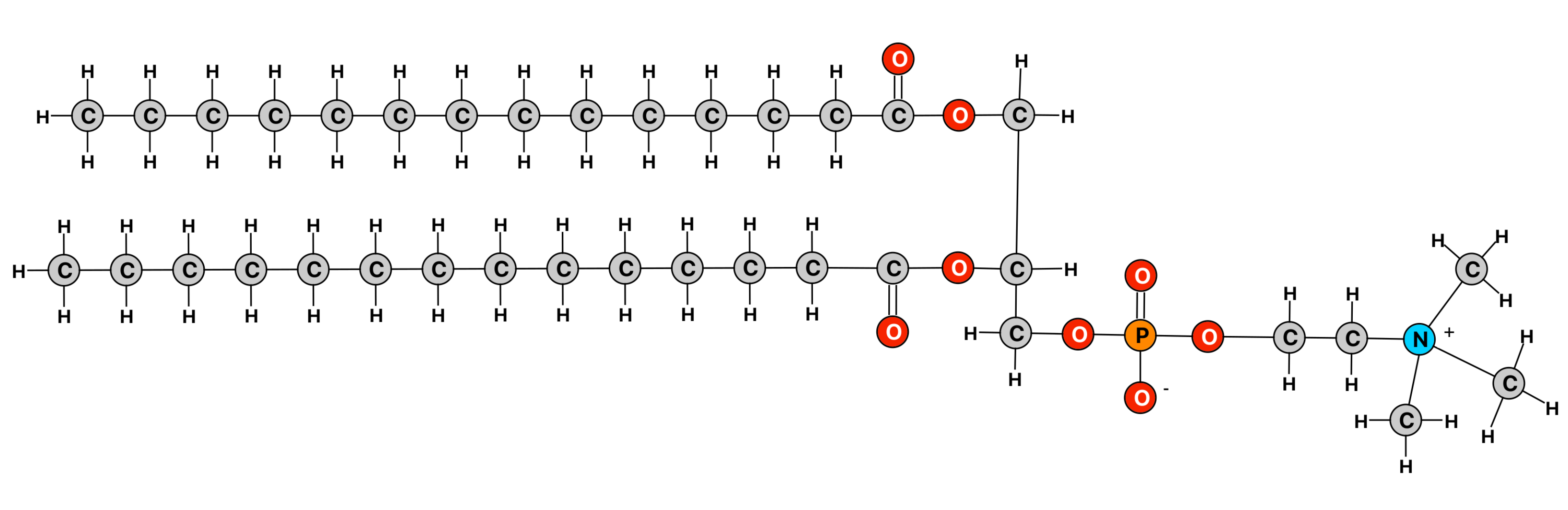}
    \caption{\label{fig:formula} 
    Chemical structure of the DMPC phospholipid.}
  \end{center}
\end{figure}

In this article, we investigate the properties of water confined by phospholipid membranes. Specifically, we measure the extent to which phospholipid membranes affect the structural properties of water as well as its HBN. As a typical model membrane, we use  1,2-Dimyristoyl-sn-glycero-3-phosphocholine (DMPC) lipids. The DMPC is a phospholipid with a choline headgroup and a tailgroup formed of two myristoyl chains(see fig.~\ref{fig:formula}). Choline-based phospholipids are ubiquitous in cell membranes and commonly used in drug-targeting liposomes~\cite{hamley}. 
In this investigation, we probe the structural properties of water by inspecting the IRO using a sensitive local metric, recently introduced by Martelli \textit{et al.} ~\cite{martelli_LOM} and already applied in a wide variety of studies~\cite{martelli_LOM,martelli_unravelling_2019,martelli_fop,martelli_chapter,martelli_searching,santra_bnnt}. Next, we examine the topology and the quality of the HBN and explore correlations of these observables with the behaviour of the structural order. We probe the topology of the HBN \textit{via} ring statistics, a tool widely adopted in network-forming materials as a measure of closed loops present in the network. Ring statistics have been previously employed in water to characterize its different phases~\cite{martelli_LOM,martelli_searching,martonak_2004,martonak_2005,camisasca_proposal}, as well as to investigate the origin of water anomalies~\cite{martelli_unravelling_2019,russo_2014,mio_nature,santra_2015,hassanali_2020}.


\section*{Results}\label{results}
\begin{figure}
  \begin{center}
   \includegraphics[scale=.50]{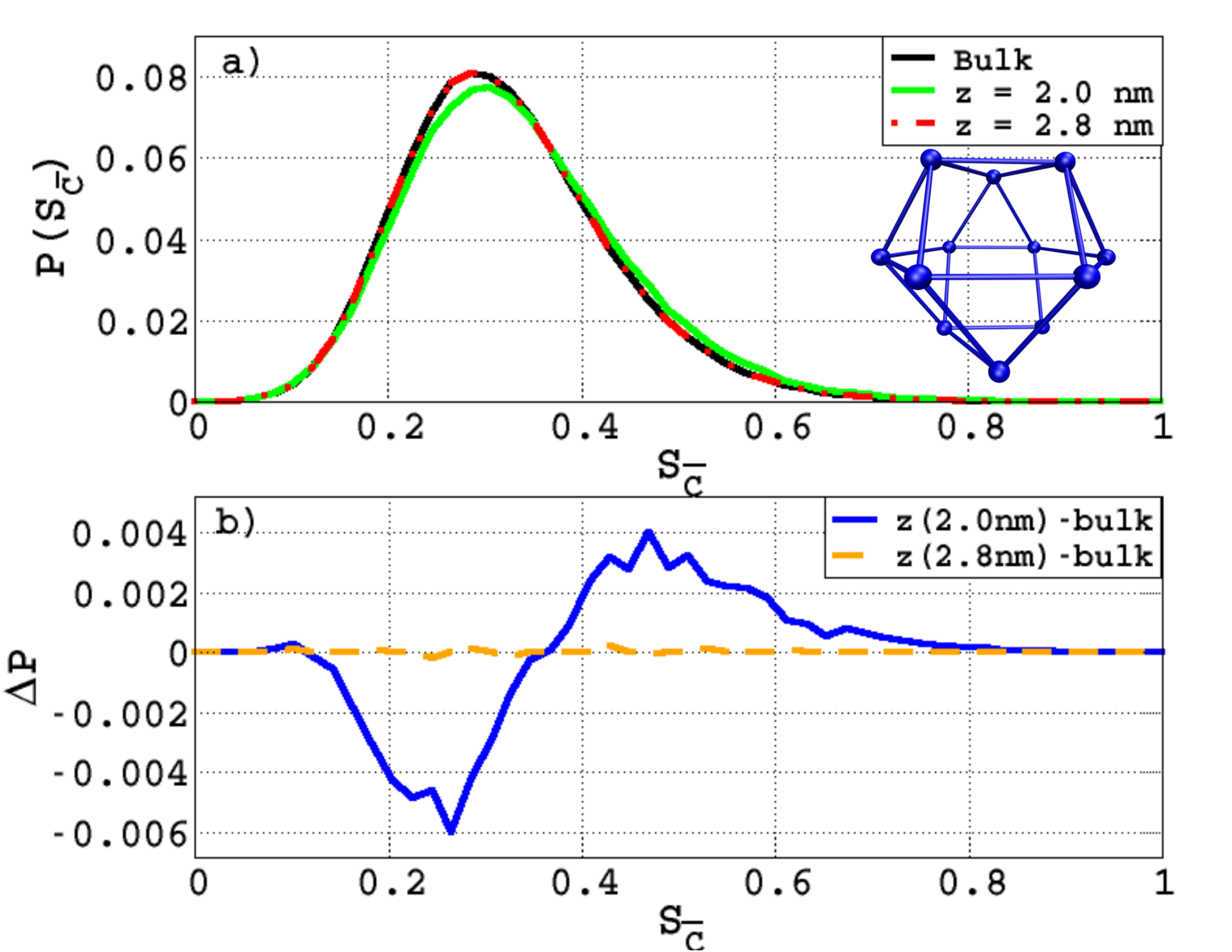}
    \caption{\label{fig:score} 
    Panel a): Probability distributions of score function $S_{\bar{C}}$ computed using the reference in the inset: the black line  is for bulk water, the green line is for water in the bin centered at $z=2.0$~nm,  the red dashed line is for water in the bin centered at $z=2.8$~nm.
    Inset: Reference made of the oxygen positions (blue spheres) of the second coordination shell in cubic ice. Blue sticks are a guide to the eyes to emphasize the geometrical structure. Panel b): Difference $\Delta P$ between $P(S_{\bar{C}})$ for bulk water  and for the bin at $z=2.0$~nm  (blue, continuous line), and between  bulk water 
     and the bin at $z=2.8$~nm   (orange, dashed line).}
  \end{center}
\end{figure}

We inspect the IRO by computing the score function (Eq.~\ref{eq:Eq3}), which provides a nonlinear measure of the deviation in atomic or molecular arrangements from those of a reference structure that, usually, corresponds to the medium's ground state at $T=0$~K. Therefore, we here compute $S$ using, as a reference, the position of the oxygen in the second coordination shell in cubic ice, \textit{i.e.}, a cuboctahedron ($\bar{C}$) that belongs to the class of Archimedean solids enriched with edge transitivity~\cite{atlas_2016} (inset in fig.~\ref{fig:score}). Similar results hold if we use, as a reference, the position of the oxygen in the second coordination shell in hexagonal ice. 

At fixed $T=303$~K and $P=1$~atm, we compare the distributions of $S_{\bar{C}}$
for  bulk water  and  confined water at a distance of $1.6$~nm $\leq z\leq 2.4$~nm  (bin centered at $2.0$~nm)
and $2.4$~nm $\leq z\leq 3.2$~nm (bin centered at  $2.8$~nm)
from the membrane (fig.~\ref{fig:score} a). We emphasize that the bin width of $0.8$~nm adopted in this work ensures that water molecules centered in the middle of the bin have a second shell of neighbours falling inside the same bin.

The distribution at $2.0$~nm does not match that of bulk water.
In particular, we find that water $2.0$~nm  away from the membrane is more structured than bulk  water, with a $\sim8\%$ increase of the $S_{\bar{C}}$ with maximum probability, and a higher population in the large-$S_{\bar{C}}$ tail of the probability distribution $P(S_{\bar{C}})$ (fig.~\ref{fig:score} b).

On the other hand, the $P(S_{\bar{C}})$ for water at $2.8$~nm from the membrane overlaps with that of bulk water, within our resolution, showing that the value of the local order metric (LOM) of water between $2.4$~nm and $3.2$~nm is not affected by the membrane. Hence, our result implies that  the effect of the membrane on the structural properties of water should extend as far as $1.6$~nm plus the second coordination shell distance ($\sim 0.45$~nm \cite{Mark:2001aa}), and at about $2.4$~nm minus $0.45$~nm, \textit{i.e.},  up to $(2.0\pm 0.05)$~nm, approximately (fig.~\ref{fig:score} b).


Since the properties of water emanate from the underlying network of HBs~\cite{martelli_unravelling_2019}, a natural question follows: Is there a connection between i) the observed perturbations on the IRO of confined water, ii) the underlying HBN, and iii) its quality in terms of broken and intact HBs? We will address this question in the following discussion.

\begin{figure}
  \begin{center}
   \includegraphics[scale=.50]{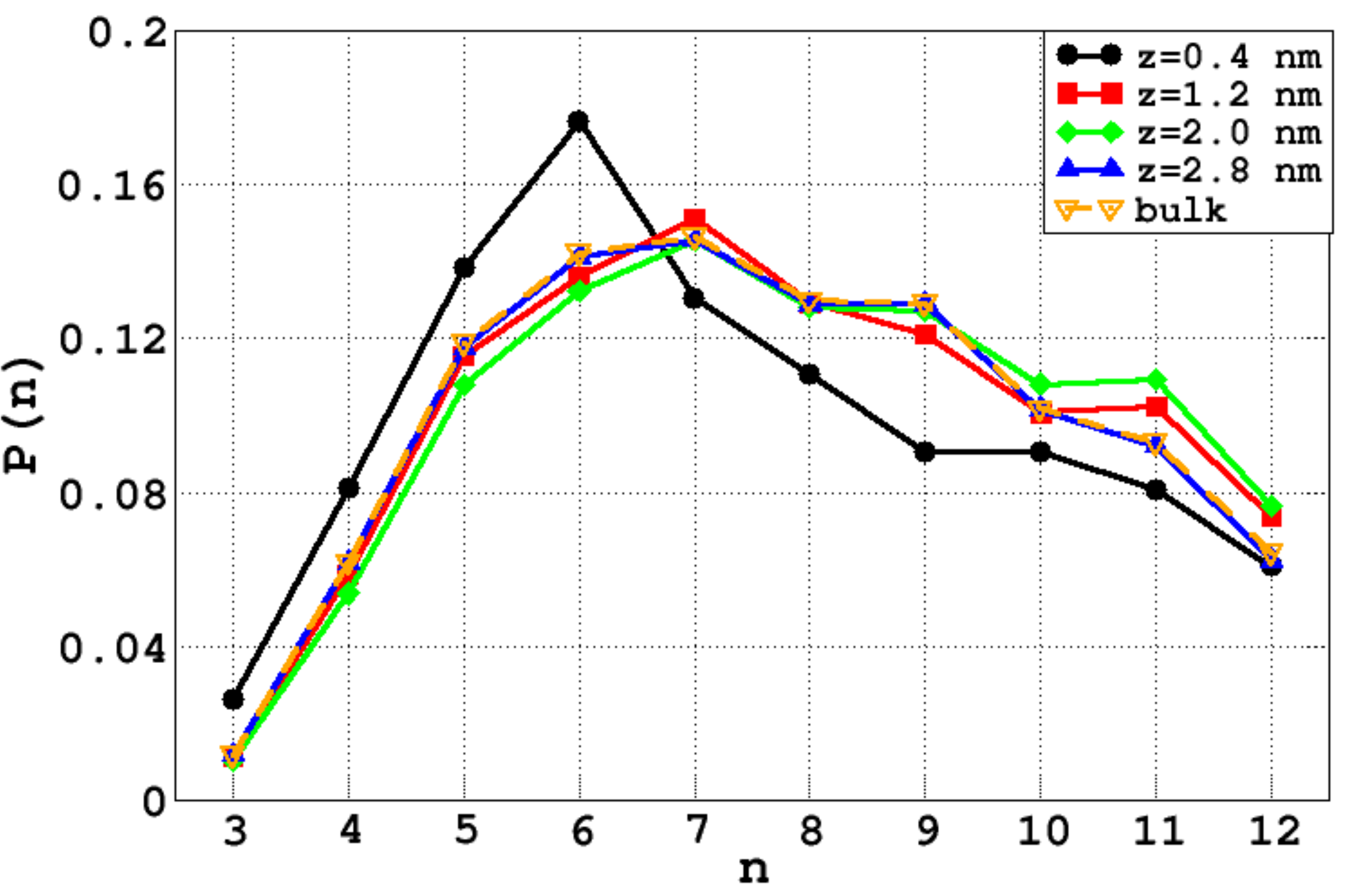}
    \caption{\label{fig:rings} Probability of the HB $n$-member rings, $P(n)$, computed from structures in bulk water (open orange triangles), and in water in bins centered at different distances from the membrane surfaces: $0.4$~nm (black dots), $1.2$~nm (red squares), $2.0$~nm (green diamonds), and $2.8$~nm (blue triangles). All $P(n)$ are normalized to unity and, therefore, do not reflect the total number of rings of a given size. }
  \end{center}
\end{figure}

We probe the HBN of water using the ring statistics, and we inspect the quality of the network quantifying and characterizing coordination defects. 
We compare the 
probability $P(n)$ of having an $n$-member ring, $n\in\left[ 3,12\right ]$, for the four bins that discretize the simulation box, and for bulk water at the same thermodynamic conditions, $T=303$~K and $P=1$~atm (fig.~\ref{fig:rings}). 

For bulk water, as expected in diffusive media, $P(n)$ is broad and accounts  for very large rings. 
We find that for distances within the bin closer to the bilayer,
\textit{i.e.}, at a distance $z\leq 0.8$~nm from the membrane, $P(n)$
strongly deviates from the corresponding probability in bulk water. 
Namely, we observe a depletion in the number of larger rings and an increase, notably sharp in the case of $n=6$, of shorter rings. We attribute the depletion of larger rings to the proximity of the membrane, which represents a reduction of dimensionality in the connectivity search pathways of water molecules. 
On the other hand, we remark that $n=6$ represents the typical connectivity in crystalline ice. 
Therefore, the increased number of hexagonal rings at $z=0.8$~nm indicates that, closer to the interface, the HBN seems to acquire a topology closer to that of an ordered crystalline network. However, as we will discuss later, this similarity is only apparent. 

The increased number of short rings at $z\leq 0.8$~nm from the membrane is in agreement with the  dynamical slowing down 
\cite{calero_membranes_2019}, and the increment in the IRO~\cite{martelli_fop} reported for similar distances.
Hence, 
1) the diffusion and rotational slowing down \cite{calero_membranes_2019},
2) the increased value of $S_{\bar{C}}$ (fig.~\ref{fig:score}), and  
3) the increased fraction of hexagonal rings (fig.~\ref{fig:rings})
for water at $z\leq 0.8$~nm from the membrane, suggest a connection between dynamics and structure as measured by 
i) the positions of the oxygen atoms and, 
ii) the topology of the HBN.

Moving from $z\leq 0.8$~nm  to  $0.8$~nm $< z\leq 2.4$~nm, 
we observe a marked change in the distribution of $n$-rings with a decrease for $n\leqslant6$ and an increase  for $n>6$ (fig.~\ref{fig:rings}). 
We attribute the larger probability for extended rings for $z>0.8$~nm to the increased dimensionality of the space available. 
In particular, 
the difference in the $P(n)$ and in the $S_{\bar{C}}$ (fig.~\ref{fig:score}), between bulk and water within the bin centered at $z=2.0$~nm from the membrane, further points toward a close correlation between structural properties at the level of the medium range, and the topology of the HBN.

The drastic change in the ring probability between the bin centered in $z=0.4$~nm and the bins at a larger distance, is consistent with the recent discovery of an interface between bound and unbound hydration water  at about $0.5$~nm  from the membrane\cite{calero_membranes_2019}. In Ref.~\cite{calero_membranes_2019} Calero and Franzese identify the interface between i) 
the first hydration shell, partially made of  water bound to the membrane, with a structural role and an extremely slow dynamics, and 
ii) 
 the next shells with no water-lipids HBs and a dynamics ten time faster than bound water, but still one order of magnitude slower than bulk water. 
Therefore, ring probability can mark the structural difference between bound and unbound water.
 
Moving to a distance $2.4$~nm $< z\leq 3.2$~nm from the surface, the $P(n)$ overlaps perfectly with the bulk case. Moreover, the $P(S_{\bar{C}})$ computed within this bin (fig.~\ref{fig:score}) overlaps with the $P(S_{\bar{C}})$ of bulk water. Therefore, we  conclude that water recovers the structural (both IRO and HBN) properties of bulk water only if at a distance larger than $2.4$~nm from the membrane. This value is 
twice the $1.2$~nm at which water retrieves bulk density and dynamics~\cite{martelli_fop,martelli_chapter}. 

To get further insights into the network topology, we inspect its quality. When water is in the glass state, we can map its HBN to a nearly-hyperuniform network, \textit{i.e.}, to a continuous random network characterized by a low fraction of coordination defects and a suppression of long-range density fluctuations~\cite{martelli_hyperuniformity}. Therefore, the number of broken HBs is a measure of the quality of the HBN. In particular, it quantifies how far the HBN is from the two extreme cases: a) the liquid and b) the continuous random network.  Furthermore, coordination defects directly affect the fluidity of liquid water. Therefore, they can be related to water dynamics \cite{delosSantos2012}.

We perform a decomposition of the HBs per water molecule into acceptor-(A) and donor-(D) types.
We label as $\textit{A}_2\textit{D}_2$ a water molecule with perfect coordination, \textit{i.e.}, donating two bonds and accepting two bonds.
We evaluate the quality of the HBN by computing the ratio of water molecules that have different coordination, \textit{i.e.},  are not in the $\textit{A}_2\textit{D}_2$ configuration. 
In particular, we focus our attention on the following coordination configurations: $\textit{A}_1\textit{D}_1$, $\textit{A}_2\textit{D}_1$, $\textit{A}_1\textit{D}_2$, $\textit{A}_2\textit{D}_2$ and $\textit{A}_3\textit{D}_2$. Other configurations do not contribute significantly~\cite{distasio_2014}.

\begin{figure}
  \begin{center}
   \includegraphics[scale=.45]{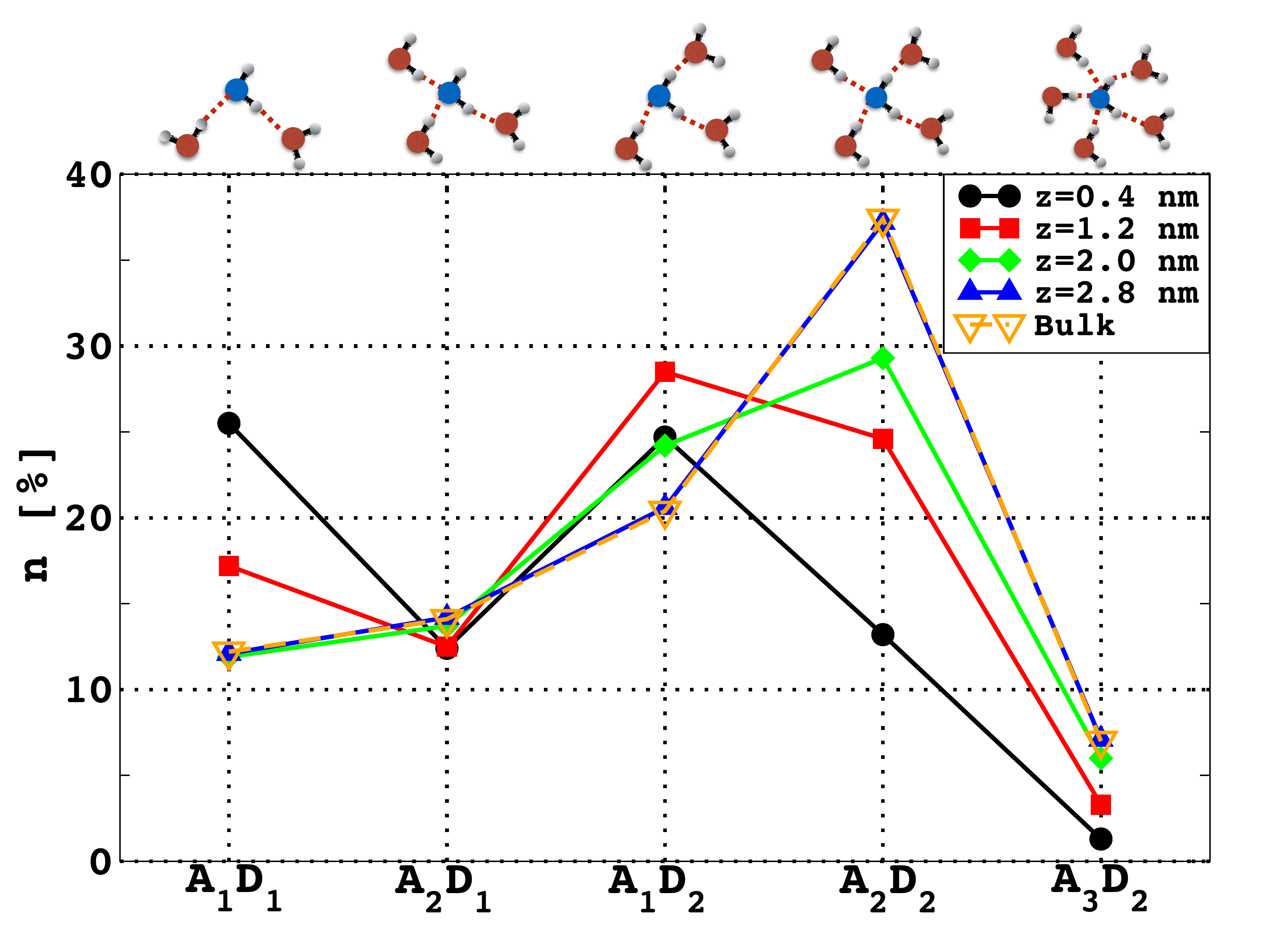}
    \caption{\label{fig:defects} 
Percentage-wise decomposition of the intact HBs per water molecule
into acceptor-(A) and donor-(D) for water in bins centered at different distances from the membrane and 
for bulk water.  
Sets are for bins at $0.4$~nm (black dots), $1.2$~nm (red squares), $2.0$~nm (green diamonds),  $2.8$~nm (blue triangles)  and bulk (orange open triangles).
The $x$-axis labels $\textit{A}_x\textit{D}_y$ indicate the number of acceptor 
($\textit{A}_x$) and donor 
($\textit{D}_y$) HBs, respectively, of the configurations schematically represented on the panel's top (with the oxygen of central water molecule in blue). For clarity we omit combinations  with minor contributions, \textit{e.g.}, 
$\textit{A}_3\textit{D}_1$, 
$\textit{A}_0\textit{D}_y$, 
$\textit{A}_x\textit{D}_0$, 
\textit{etc.}}
  \end{center}
\end{figure}

We compare the percentage of intact HBs for bulk and confined water, as a function of the distance from the membrane (fig. \ref{fig:defects}). 
 We find that the HBN in bulk water is dominated by $\textit{A}_2\textit{D}_2$ ($\sim37\%$) perfect coordinations. Water molecules involved in three HBs in the form of the defect $\textit{A}_1\textit{D}_2$ comprise the next largest percentage ($\sim20\%$), followed by the $\textit{A}_2\textit{D}_1$ and $\textit{A}_1\textit{D}_1$ types and, finally, by the $\textit{A}_3\textit{D}_2$. 
 
This result, based on TIP3P water, is in agreement with the trend in \emph{ab initio} liquid water at ambient conditions examined with different functionals~\cite{distasio_2014}. In particular, in \emph{ab initio} liquid water, the frequency of $\textit{A}_1\textit{D}_2$ is almost twice that of $\textit{A}_2\textit{D}_1$ at all levels of theories~\cite{distasio_2014}.
 
Close to the surface of the membrane, at $z \leq 0.8$~nm, the network of HBs largely deviates from that of bulk water. The network is dominated by $\textit{A}_1\textit{D}_1$ and $\textit{A}_1\textit{D}_2$ defects ($\sim25\%$), followed by $\textit{A}_2\textit{D}_1$ and $\textit{A}_2\textit{D}_2$ configurations ($\sim15\%$), and a small percentage of higher coordination defects $\textit{A}_3\textit{D}_2$ ($\sim3\%$). Such composition is very consistent with the results found for bound water at $z \leq 0.5$~nm \cite{calero_membranes_2019}. 
In particular, we find here the same percentage of defects with three water-water HBs ($40\%$). Furthermore, we observe  numbers, very close to those in Ref.\cite{calero_membranes_2019}\ ,
 for perfectly coordinated configurations ($\sim 20\%$), and defects with two ($\sim 30\%$) and five water-water HBs ($\sim 1\%$).

However, close to the membrane, the decrease of perfectly coordinated water molecules 
seems to be inconsistent with the higher local order of water~\cite{martelli_fop,martelli_chapter}, and also with the enhanced contribution of six-fold rings (fig.~\ref{fig:rings}). This discrepancy is only apparent, for two reasons. First,  
 both the IRO and the ring statistics are a measure of local order beyond short range, while the quality of the HBN is strictly a short-range measure. 
Second, our calculations include only water-water HBs and do not account for the (strong) HBs between water molecules and the phospholipid headgroups~\cite{binder_2003,chen_2010,calero_membranes_2019}. Instead, $\sim30\%$ of the water molecules in the first hydration shell are bound to the membrane with at least one HB \cite{calero_membranes_2019}. This observation explains why 
 the dynamical slowing down ~\cite{martelli_fop} of bound water can be interpreted as a local reduction of thermal noise that allows water molecules to organize in space in more ordered geometrical configurations \cite{calero_membranes_2019}.

Moving away from the surface, at a distance of $0.8$~nm$<z\leq 2.4$~nm, the most appreciable effect on the quality of the HBN is a marked reduction of $\textit{A}_1\textit{D}_1$ defects down to $\sim18\%$, mostly accounting for the absence of HBs between water molecules and phospholipid headgroups \cite{calero_membranes_2019}, and a corresponding drastic increase in the percentage of perfectly coordinated water molecules ($\textit{A}_2\textit{D}_2$) up to $\sim25\%$, confirming the analysis done for unbound water  \cite{calero_membranes_2019}.

At these distances bulk density and dynamical properties of water are recovered almost fully~\cite{martelli_fop, calero_membranes_2019}. However, the quality (defects) of the HBN strongly deviates from that of bulk water, accounting for its different topology (ring probability, fig.~\ref{fig:rings}) and its different structural properties ~\cite{martelli_fop}. 

Upon increasing the distance from the membrane, we find a reduction of most of the coordination defects and a corresponding increase of perfectly coordinated water molecules, \textit{i.e.}, an improvement in the quality of the HBN (fig. \ref{fig:defects}). Nevertheless, we recover
 the bulk-like composition only at a distance $z>2.4$~nm from the membrane, as in our analysis for both the IRO (fig.~\ref{fig:score}) and the topology of the HBN (fig.~\ref{fig:rings}).

It is interesting to observe that the percentage of the defect type $\textit{A}_2\textit{D}_1$ is mostly constant in all bins and, therefore, is independent of the distance from the membrane (fig.~\ref{fig:defects}). We are currently working on rationalizing this intriguing evidence.

\section*{Conclusions}\label{conclusions}

The relation between water dynamics and structure is elusive in bulk \cite{Verde:2019aa} and even more 
under nanoconfinement \cite{Joseph:2008dz}, especially when the confining surfaces are soft  \cite{Ruiz-Pestana:2018aa}. On the other hand, the relationships among the hydration structure and molecular fluidity at  membrane/water interfaces are relevant in many biological processes \cite{Asakawa:2012aa}.
Here,
we study why water recovers its density and dynamics at $\sim1.2$~nm from a membrane~\cite{martelli_fop, calero_membranes_2019} while has an intermediate range order (IRO)~\cite{martelli_LOM} higher than bulk up to a distance twice as large~\cite{martelli_fop}.
To understand this surprising result,
we focus on the hydrogen bond network (HBN), analyzing its topology (ring statistics) and its quality (population of perfectly coordinated water molecules and defects).
We find that the increased IRO is characterized by an alteration of the HBN.

In particular, for bound water~\cite{calero_membranes_2019}, \textit{i.e.},  water  at short distances (here less than $0.8$~nm) from the membrane, we show that the HBN topology and quality are very different from those of low-temperature bulk water. Although bound water has an HBN with a large fraction of hexagonal rings as in crystalline water, it has a much higher number of defects than low-temperature water. 
We find that  $\textit{A}_1\textit{D}_1$ and  $\textit{A}_1\textit{D}_2$ account together for 50\% of all the defects
due to water strong HBs with the membrane. 
These strong HBs locally reduce the water's thermal energy and slow down its dynamics.

We show that the HBN analysis is able to mark the existence of the bound-unbound water interface~\cite{calero_membranes_2019}.  
We find a sudden qualitative change in the ring statistics for hydration water at a distance $z>0.8$~nm from the membrane. 
Also the defects distribution clearly shows that water in the range  $0.8$~nm$<z\leq 2.4$~nm is neither bound to the membrane, neither bulk. 
Indeed, it has  much less  $\textit{A}_1\textit{D}_1$ defects than bound water,  
and much less perfectly-coordinated molecules than bulk. 
Nevertheless, at these distances, the structural differences between unbound and bulk water are disguised in water's density and dynamics~\cite{calero_membranes_2019}.
 
The difference in topology and defects smear out at distances larger than $2.4$~nm. 
This distance  
corresponds to more than eight coordination shells of hydration water. Hence, our results
support the evidence of long-range effects measured in terahertz and dielectric relaxation experiments~\cite{havenith_2007,Tielrooij_BiophysJ2009,cui_2013,hishida_long_2013}. 
We expect our conclusions to hold and eventually be emphasized by water potentials more realistic than TIP3P, which is quite poor in terms of structural properties beyond the short-range.

Our findings should be taken into account when interpreting experimental results and when developing membrane-water interaction potentials. They can help in better understanding water in biological processes at large, in particular those where hydration or structural changes play a role. Variations of ions concentration drastically change the water HBN~\cite{Mancinelli:2007fk} and its dynamics~\cite{Fayer:2009zx}, with an effect that is similar to an increase of pressure~\cite{Gallo:2014ab}, or a decrease of temperature for dehydration~\cite{calero_membranes_2019}. These variations in the extracellular matrix can promote, for example, cardiac disease and arterial hardening in healthy men~\cite{Arnaoutis:2017aa} or atherosclerosis and inflammatory signaling in endothelial cells~\cite{10.1371/journal.pone.0128870}. 
Hence, our results entail further investigation about the relationship between this category of diseases with the water HBN rearrangements due to changes in hydration or ionic concentrations.

\section*{Methods}\label{methods}
\subsection*{Simulation details}
\begin{figure}
  \begin{center}
   \includegraphics[scale=.45]{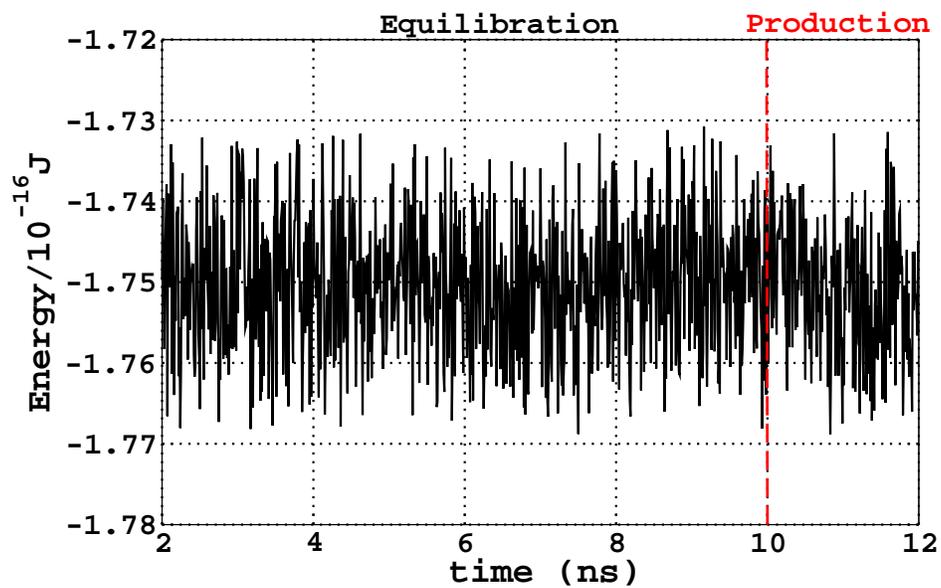}
    \caption{\label{fig:energy} 
     Energy profile for the system under consideration. The red dashed arrow defines the limit after which we start the production and data analysis.}
  \end{center}
\end{figure}

The systems considered here have the same geometry as in our previous simulations~\cite{martelli_fop} but with a 15\% increase in hydration, \textit{i.e.}, they are composed of 128 DMPC lipids in a bilayer and $8100$ water molecules, with periodic boundary conditions in such a way that water is confined between the two sides of two replicas of the same membrane. 
We perform molecular dynamics (MD) simulations on IBM POWER8 machines with NVIDIA Kepler K80 GPUs using the simulation package NAMD 2.9~\cite{Phillips_JCompChem2005} at a temperature of $T=303$~K and an average pressure of $p=1$~atm. We set the simulation timestep to $2$~fs. We describe the structure of phospholipids and their mutual interactions by the recently parameterized force field CHARMM36~\cite{Klauda_JPhysChemB2010,charmm_2}, which is able to reproduce the area per lipid in excellent agreement with experimental data. The water model employed in our simulations, consistent with the parametrization of CHARMM36, is the modified TIP3P~\cite{tip3p_1}. We cut off the Van der Waals interactions at $12$~\AA\  with a smooth switching function starting at $10$~\AA. We compute the long-ranged electrostatic forces with the particle-mesh Ewald method~\cite{Essmann_JCP1995}, using a grid spacing of $1$~\AA. 
Our simulation box is anisotropic, with $L_z > L_x$, with $L_x=L_y$. This anisotropy ensures that there are no errors caused by the calculation of long-range electrostatics. During the $NpT$ simulations we always keep this condition, with $\sim5\%$ fluctuations for the values of $L_x$, $L_y$, and $L_z$.
After energy minimization, we equilibrate the hydrated phospholipid bilayers for $10$~ns followed by a production run of $2$~ns in the $NpT$ ensemble at $p=1$~atm. The energy profile is shown in fig.~\ref{fig:energy}.

In the simulations, we control the temperature with a Langevin thermostat~\cite{Berendsen_JPhysChem1984} using a damping coefficient of $0.1$~ps$^{-1}$ and we control the pressure by a Nos\'{e}-Hoover Langevin barostat~\cite{Feller_JChemPhys1995} with a piston oscillation time of $0.2$~ps and a damping time of $0.1$~ps. We also perform numerical simulations of bulk TIP3P water (4000 molecules) adopting the same protocol and at the same thermodynamic conditions. It is worthy to mention, at this point, that the isotropic $NpT$ ensures that experimental observables such as, \textit{e.g.}, the area per lipid and NMR order parameters, are properly reproduced~\cite{taylor_2009,davis_2009,gapsys_2013}.

In order to investigate the IRO and the HBN, we divide the systems along the direction perpendicular to the phospholipid bilayer in equally spaced bins such that the thickness of each bin is $0.8$~nm.
This thickness is chosen in such a way that a molecule of water in the bin's center has, approximately, it's second coordination shell included in the same bin at $T=303$~K and $P=1$~atm.
For our hydration and bin's size, the two membranes are separated by eight bins, hence we can analyze four different distances between 0 and $3.2$~nm.
All the distances are measured taking as reference distance, for each side of the membrane, the position where the phospholipid density distribution, at thermodynamic equilibrium, has a maximum. 
We measure the observables of interest for each water molecule within each bin centered at a distance $z$ from the center of the bilayer. It is worthy to mention that several more sophisticated ways of computing the distances from a rough membrane surface have been reported in the literature~\cite{calero_membranes_2019,pandit_algorithm}. On the other hand, such methods show differences whit respect to our approach when thinner bins are implemented and only in the proximity of the surface.

\subsection*{The local order metric}
We here briefly discuss the basic ideas behind the LOM we have used to probe the structural properties of water. Details can be found in Ref.~\cite{martelli_LOM}.

The local environment of a water molecule $j$ in a snapshot defines a local {\it pattern} formed by $M$ neighboring sites.
Here, we consider only the oxygen atoms' second neighbors of the oxygen of the molecule $j$.

There are $N$ local patterns,
one for each atomic site $j$ in the system. Indicating by $\mathbf{P}_{i}^{j} (i=1,M)$ the position vectors
in the laboratory frame of the $M$ neighbors of site $j$, their centroid is given by
$\mathbf{P}_{c}^{j}\equiv\frac{1}{M}\sum_{i=1}^{M}\mathbf{P}_{i}^{j}$.
In the following we refer the positions of the sites of the pattern to their centroid, \textit{i.e.}
$\mathbf{P}_{i}^{j}-\mathbf{P}_{c}^{j}\rightarrow \mathbf{P}_{i}^{j}$. 

The local {\it reference} is a set of
$M$  sites, labeled by indices $i (i=1,M)$, located at positions $\mathbf{R}_{i}^{j}$ around the molecules $j$ 
in ideal positions, typically as in a lattice of choice. 
The step
of the reference lattice is fixed equal to equilibrium O--O distance, $d$,  in the water coordination shell at the  thermodynamic conditions of interest.

For each oxygen site $j$ the centroid of the reference is set to coincide with the centroid of the pattern. The reference orientation is, instead, arbitrary, forming angles $\theta,\phi,\psi$ with the pattern. 

The LOM $S(j)$ at site $j$ is the maximum of the overlap function
with respect to the orientation of the reference and the permutation of the pattern
indices, 
\begin{equation}
  S(j)\equiv \max_{\theta,\phi,\psi;\mathcal{P}}\prod_{i=1}^{M}\exp\left(-\frac{\left| \mathbf{P}_{i_{\mathcal{P}}}^{j}-\mathbf{R}_{i}^{j}\right|^2}{2\sigma^{2}M}\right).
  \label{eq:Eq1}
\end{equation}
Here 
$i_{\mathcal{P}}$ are the permuted indices of the pattern sites corresponding to a permutation $\mathcal{P}$,
and $\sigma=d/4.4$ is a parameter that controls the spread of the Gaussian functions.

If $L$ is the number of proper point symmetry operations of the reference,
the overlap function (Eq.~\ref{eq:Eq1})  has $L$ equivalent maxima. Therefore,
it is sufficient to compute $S(j)$ for 
only a fraction $1/L$ of the Euler angle domain $\Omega$,
which we may call $\Omega/L$, the irreducible domain of the Euler angles.
Inside $\Omega/L$ we pick
at random, with uniform probability, $15$ orientations and we optimize them using a conjugate gradients procedure. \par
The LOM is an intrinsic property of the local environment at variance with the overlap
function $\mathcal{O}(j)$ that depends on the orientation of the reference and on the ordering of the
sites in the pattern. The LOM satisfies the inequalities
$0 \lesssim S(j) \leq 1$. The two limits correspond, respectively, to a completely disordered local
pattern ($S(j)\rightarrow 0$) and to an ordered local pattern matching perfectly the reference ($S(j)\rightarrow 1$),
therefore grading each
local environment on an increasing scale of local order from zero to one. \par
The order parameter {\it score function} $S$  is the site-averaged LOM:
\begin{equation}
  S\equiv \frac{1}{N}\sum_{j=1}^{N}S(j),
  \label{eq:Eq3}
\end{equation}

\subsection*{Definition of rings}
Several definitions of rings and counting schemes have been reported in the literature~\cite{king,rahman_hydrogen,guttman_ring,franzblau_computation,wooten_structure,yuan_efficient,leroux_ring}. 
Recently, Formanek and Martelli have shown that different schemes allow us to access different information~\cite{martelli_rings}. 
 
\begin{figure}
  \begin{center}
   \includegraphics[scale=.75]{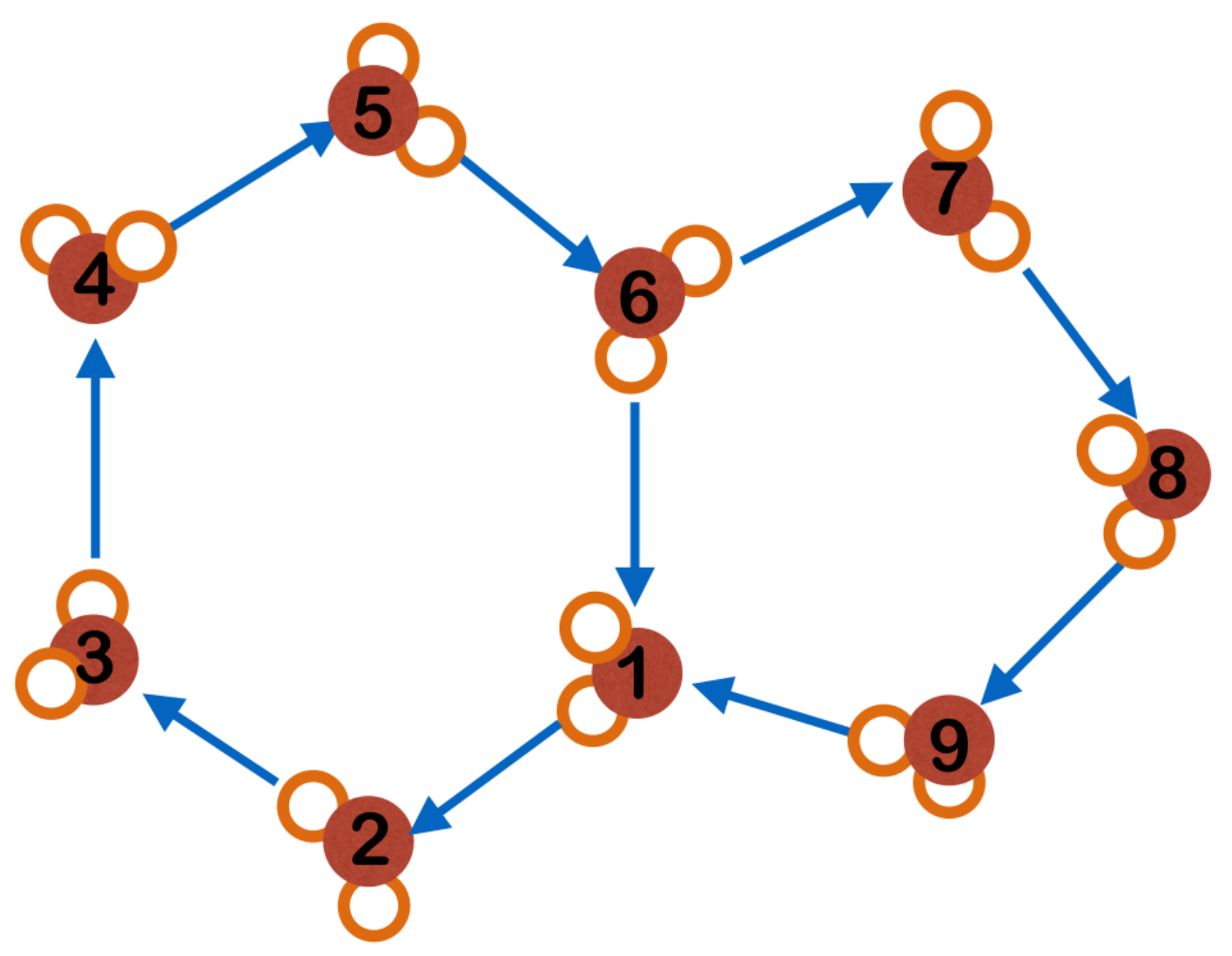}
    \caption{\label{fig:ring} Schematic representation of the ring construction for a given 
    HBN between water molecules. We start from water molecule labeled as 1 (O atoms are represented in red, H atoms in white). By following the directional HBs from H to O  (blue arrows), we  cross the HBN until we return to molecule 1 or we exceeds 12 steps and then take only those rings that cannot be decomposed in sub-rings. Here, we find a hexagonal ring  from molecule 1 to 6 and and a pentagonal ring from  1 to 9.
    }
  \end{center}
\end{figure}

Here, we construct rings as in fig.~\ref{fig:ring}.
We adopt the geometric definition of HB ~\cite{chandler_HB}, that is in qualitative agreement with other definitions over a wide range of thermodynamic conditions~\cite{prada_2013,shi_2018_2}. 
We start from a tagged water molecule and recursively traverse the HBN until we reached again the starting point, or we exceed the maximal ring size considered, 12 water molecules in our case. We consider only the primitive rings, \textit{i.e.}, rings that can not be decomposed into smaller ones~\cite{goetzke_properties,wooten_structure,king}. 
As shown in Ref.~\cite{martelli_rings}, this definition provides a rich amount of information about the network. 

\section*{Acknowledgements}
F.M. and J.C. acknowledge support from the STFC Hartree Centre's Innovation Return on Research programme, funded by the Department for Business, Energy and Industrial Strategy. 
G.F. acknowledges support from the Spanish grant PGC2018-099277-B-C22 
(MCIU/ AEI/ ERDF) and ICREA Foundation (ICREA Academia prize).

\bibliographystyle{unsrt}
\bibliography{main}

\end{document}